\documentclass[12pt,preprint]{aastex}
\usepackage{booktabs}
\begin{document}

\title{OGLE-2012-BLG-0724Lb: A Saturn-mass Planet around an M-dwarf.}

\author{Y. Hirao\altaffilmark{1,2}, A. Udalski\altaffilmark{3,4}, T. Sumi\altaffilmark{1,2}, D.P. Bennett\altaffilmark{5,6,2}, I.A. Bond\altaffilmark{6,2}, N. Rattenbury\altaffilmark{7,2}, D. Suzuki\altaffilmark{5,6,2}, N. Koshimoto\altaffilmark{1,2} \\ and \\F. Abe\altaffilmark{8}, Y. Asakura\altaffilmark{8}, A. Bhattacharya\altaffilmark{5}, M. Freeman\altaffilmark{9}, A. Fukui\altaffilmark{10}, Y. Itow\altaffilmark{8}, M.C.A. Li\altaffilmark{7}, C.H. Ling\altaffilmark{11}, K. Masuda\altaffilmark{8}, Y. Matsubara\altaffilmark{8}, T. Matsuo\altaffilmark{1}, Y. Muraki\altaffilmark{9}, M. Nagakane\altaffilmark{1}, K. Ohnishi\altaffilmark{12}, H. Oyokawa\altaffilmark{8}, To. Saito\altaffilmark{13}, A. Sharan\altaffilmark{7}, H. Shibai\altaffilmark{1}, D.J. Sullivan\altaffilmark{14}, P.J. Tristram\altaffilmark{15}, A. Yonehara\altaffilmark{16}, \\ (The MOA Collaboration) \\ 
R. Poleski\altaffilmark{17}, J. Skowron\altaffilmark{3}, P. Mr\'oz\altaffilmark{3}, M.K. Szyma\'nski\altaffilmark{3}, S. Koz\l owski\altaffilmark{3}, P. Pietrukowicz\altaffilmark{3}, I. Soszy\'nski\altaffilmark{3}, \L. Wyrzykowski\altaffilmark{3}, K. Ulaczyk\altaffilmark{18} \\ (The OGLE Collaboration)}

\altaffiltext{1}{Depertment of Earth and Space Science, Graduate School of Science, Osaka University, 1-1 Machikaneyama, Toyonaka, Osaka 560-0043, Japan}
\altaffiltext{2}{Microlensing Observations in Astrophysics (MOA)}
\altaffiltext{3}{Warsaw University Observatory, A1. Ujazdowski 4, 00-478 Warszawa, Poland}
\altaffiltext{4}{Optical Gravitational Lens Experiment (OGLE)}
\altaffiltext{5}{Department of Physics, University of Notre Dame, Norte Dame, IN 46556, USA}
\altaffiltext{6}{Laboratory for Exoplanets and Stellar Astrophysics, NASA/Goddard Space Flight Center, Greenbelt, MD 20771, USA}
\altaffiltext{7}{Department of Physics, University of Auckland, Private Bag 92019, Auckland, New Zealand}
\altaffiltext{8}{Institute for Space-Earth Environmental Research, Nagoya University, Nagoya 464-8601, Japan}
\altaffiltext{9}{Department of Physics, University of Auckland, Private Bag 92019, Auckland, New Zealand}
\altaffiltext{10}{Okayama Astrophysical Observatory, National Astronomical Observatory of Japan, 3037-5 Honjo, Kamogata, Asakuchi, Okayama 719-0232, Japan}
\altaffiltext{11}{Institute of Information and Mathematical Sciences, Massey University, Private Bag 102-904, North Shore Mail Centre, Auckland, New Zealand}
\altaffiltext{12}{Nagano National College of Technology, Nagano 381-8550, Japan}
\altaffiltext{13}{Tokyo Metropolitan College of Aeronautics, Tokyo 116-8523, Japan}
\altaffiltext{14}{School of Chemical and Physical Sciences, Victoria University, Wellington, New Zealand}
\altaffiltext{15}{Mt. John University Observatory, P.O. Box 56, Lake Tekapo 8770, New Zealand}
\altaffiltext{16}{Department of Physics, Faculty of Science, Kyoto Sangyo University, Kyoto 603-8555, Japan}
\altaffiltext{17}{Department of Astronomy, Ohio State University, 140 W. 18th Ave., Columbus, OH 43210, USA}
\altaffiltext{18}{Department of Physics, University of Warwick, Gibbet Hill Road, Coventry, CV4 7AL, UK}

\begin{abstract}
We report the discovery of a planet by the microlensing method, OGLE-2012-BLG-0724Lb. Although the duration of the planetary signal for this event was one of the shortest seen for a planetary event, the anomaly was well covered thanks to high cadence observations taken by the survey groups OGLE and MOA. By analyzing the light curve, this planetary system is found to have a mass ratio $q=(1.58\pm0.15)\times10^{-3}$. By conducting a Bayesian analysis, we estimate that the host star is an M-dwarf star with a mass of $M_{\rm L}=0.29_{-0.16}^{+0.33} \ M_{\sun}$ located at 
$D_{\rm L}=6.7_{-1.2}^{+1.1} \ {\rm kpc}$ away from the Earth and the companion's mass is $m_{\rm P}=0.47_{-0.26}^{+0.54} \ M_{\rm Jup}$. The projected planet-host separation is  $a_{\perp}=1.6_{-0.3}^{+0.4} \ {\rm AU}$. Because the lens-source relative proper motion is relatively high, future high resolution images would detect the lens host star and determine the lens properties uniquely. This system is likely a Saturn-mass exoplanet around an M-dwarf and such systems are commonly detected by gravitational microlensing. This adds an another example of a possible pileup of sub-Jupiters $(0.2 < m_{\rm P}/M_{\rm Jup} < 1)$ in contrast to a lack of Jupiters ($\sim 1 - 2 \ M_{\rm Jup}$) around M-dwarfs, supporting the prediction by core accretion models that Jupiter-mass or more massive planets are unlikely to form around M-dwarfs.
\end{abstract}


\section{INTRODUCTION}
Since the first discovery of exoplanets \citep{may95}, about 1900 exoplanets have been found  to date. They include a wide variety of planetary systems such as hot Jupiters and eccentric planets etc. But a comprehensive planetary formation model that explains all these planet types has not yet been established.

Most exoplanets have been found by the radial velocity \citep{but06} and transit \citep{bor11} methods. These methods are sensitive to low-mass planets orbiting less than 1 AU from the host star. In contrast, gravitational microlensing is sensitive to planets orbiting at a few AU away from the host stars down to Earth mass, which is complementary to other methods \citep{ben96}. \citet{sum10} constructed the planet mass function beyond the snow-line and found Neptune-mass planets are $\sim$3 times more common than Jupiter-mass planets based on 10 microlensing planets. \citet{gou10} estimated the planet abundance beyond the snow-line based on 6 events and found they are 7 times more likely than that at close orbits of 1 AU. \citet{cas12} used 6 years of PLANET collaboration data to constrain the cool planetary mass function for masses $5 M_{\oplus} - 10 M_{\rm Jup}$, and separation $0.5 - 10 AU$, including constraints obtained by the two previous works by \citet{sum10} and \citet{gou10}, and found there is one planet per star in the Galaxy. \citet{cla14} found that the planet abundances estimated by microlensing are consistent to these found by radial velocity. Recently \citet{shv15} estimated planet abundance and mass function based on 9 events and found that 55\% of stars host a planet beyond the snow-line and Neptune-mass planets are $\sim$10 times more common than Jupiter-mass planets. \citet{suz16} found a break and possible peak in the exoplanet mass ratio function at $q \sim 10^{-4}$, and found 0.75 planets per star at $q > 5 \times 10^{-5}$ and 1.12 planets per star for $q > 0$. However, the numbers of planets used in these analyses were small. Thus discovering more statistics microlensing planets is important for improving our statistical estimates of planets abundances. 

In microlensing, we can measure the planet/host mass ratio $q$ and their separation $s$ in units of Einstein radii. The mass and the distance can be measured when we detect the microlensing parallax effect or detect the lens directly together with an angular Einstein radius, $\theta_E$, obtained from the finite source effect. Also, there are other effects that can help break the degeneracy such as astrometric microlensing \citep{gou14} and interferometry \citep{cas16}. Recently, follow-up observations by {\sl Spitzer} helped resolve the degeneracy by detecting parallax effects \citep{uda15, zhu15}. Further, space-based parallax observations of microlensing events by the {\sl Kepler} telescope (K2C9) are planned in 2016 \citep{gou13}. But we need to wait for results from the {\sl WFIRST} satellite, for large number statistics. Until then, we need to collect as many ground-based microlensing events as possible, including events without parallax measurements (e.g. Suzuki et al. in preparation). In such an analysis, a careful treatment of events with a weak signal or with a degeneracy between parameters is important. 

In this paper, we report the analysis of a microlensing event OGLE-2012-BLG-0724 which has a relatively weak planetary signal and degenerate solutions. We describe the observation of this event in section 2. Section 3 explains our data reduction procedure. Section 4 discusses the modeling of the lightcurve and the comparison with other models. Section 5 discusses the likelihood analysis. Finally we discuss the results of this work in section 6.

\section{OBSERVATION}
The Microlensing Observations in Astrophysics （MOA） collaboration conducts a high cadence microlensing survey observation program toward the Galactic bulge at Mt John in New Zealand \citep{bon01,sum03}. The second phase of MOA uses the 1.8m MOA-II telescope equipped with a very wide field-of-view (2.2 ${\rm deg}^2$) MOA-cam3 CCD camera \citep{sak08}. The MOA-II observing strategy is that 6 fields ($\sim13 \ {\rm deg}^2$) with the highest lensing rate are observed with a 15 minute cadence, while the 6 next best fields are observed with a 47 minutes cadence, and 8 additional fields are observed with a 95 minutes cadence. The MOA-II observations are carried out in the custom MOA-Red wide band filter, which corresponds to the sum of the standard Cousins $R$ and $I$-bands. MOA issues $\sim 600$ alerts of microlensing events in real time each year.\footnote[1]{ https://it019909.massey.ac.nz/moa/} The 61 cm B\&C telescope at the same site is used for follow-up observations with standard $I$ and $V$-band filters. 

The Optical Gravitational Lensing Experiment (OGLE; \citep{uda15a}) conducts a microlensing survey with the 1.3 m Warsaw telescope at the Las Campanas Observatory in Chile. The fourth phase of OGLE, OGLE-IV started its high cadence survey observations in 2010 with a 1.4 ${\rm deg}^2$ FOV mosaic CCD camera. OGLE-IV observes the Galactic bulge fields with cadences ranging from one observation every 20 minutes for 3 central fields to less than one observation every night for the outer bulge fields. Most observations are taken in the standard Kron-Cousin $I$-band with occasional observations in the Johnson $V$-band. OGLE-IV issues alerts for $\sim 2000$ micolensing events in real time each year.\footnote[2]{http://ogle.astrouw.edu.pl/ogle4/ews/ews.html}

The gravitational microlensing event OGLE-2012-BLG-0724 was first found and alerted by OGLE on 22 May 2012 (HJD$'$ = HJD-245000 $\sim$6069.60) at (R.A., decl.)(J2000) = ($17^{h}55^{m}52^{s}.39$, $-29^{\circ}49'06''.7$) or $(l, b) = (0^{\circ}.385, -2^{\circ}.371)$ in Galactic coordinates. MOA independently detected and alerted the same event on the next day (HJD$'$ = 6071.02) as MOA-2012-BLG-323. The light curves are shown in Figure 1. Because the magnification of this event was predicted very high, $\sim$ 100, which is sensitive to planets \citep{gou92,rat02}, this event was alerted as a high-magnification event. During the event, the small deviation from the point-source point-lens (PSPL) model fit was detected in the MOA-II data in real-time. So follow-up observations were started using the B\&C telescope immediately and increased the observation cadence. The data covered the both caustic entry and exit, but the photometry accuracy at the caustic exit was a little worse due to thin cloud. Because the duration of the anomaly was short, about 6 hours, and the peak magnitude of $I_{\rm max} = 16.5$ was too faint for small telescopes, only the B\&C telescope could conduct follow-up observations of the anomaly.

\section{DATA REDUCTION}
The MOA-Red-band and B\&C $I$ and $V$-band data were reduced by the MOA Difference Image Analysis (DIA)  pipeline \citep{bon01}. The OGLE $V$ and $I$-band data were reduced by the OGLE DIA photometry pipeline \citep{uda15a}. Systematic photometry trends in the MOA data with seeing and differential refraction measured in years other than 2012 were linearly subtracted from all the MOA photometry data.

It is known that the nominal errors from the photometric pipeline are underestimated for the stellar dense fields like ours. We renormalized the error bars of each dataset by using the following standard formula, 
\begin{equation}
\sigma'_i = k \sqrt{\sigma_i^2 + e_{\rm min}^2}
\end{equation}
where $\sigma_i$ is the original error of the  $i$th data point in magnitudes, and $k$ and $e_{\rm min}$ are the re-normalizing parameter \citep{yee12}. The cumulative $\chi^2$ distribution from the tentative best fit model, which are sorted by their magnification of the model at each data point, is supposed to be the straight line if the data is normal distribution.
Thus $e_{\rm min}$ is chosen to make this cumulative $\chi^2$ distribution a straight line. 
Then $k$ is chosen so that each data set gives $\chi^2/{\rm d.o.f} = 1$. The results are shown in Table \ref{tab1}.

\section{LIGHT CURVE MODELING}
For a binary lens model, fitting parameters are the time of closest approach to the lens masses, $t_{\rm 0}$, $t_{\rm E}$ and $u_0$ are the Einstein radius crossing time and the impact parameter in units of the angular Einstein radius $\theta_{\rm E}$, $q$ is the planet/host mass ratio, $s$ is the planet-star separation in Einstein radius units, and $\alpha$ is the angle of the source trajectory relative to the binary lens axis. When we take account of the finite source effect and the parallax effect, the angular radius of the source star in units of $\theta_{\rm E}$, $\rho$, and the east and north components of the microlensing parallax vector, $\pi_{\rm E,E}$ and $\pi_{\rm E,N}$, are added for each case. The modeling light curve can be given by,
\begin{equation}
F(t) = A(t)F_{\rm S} + F_{\rm b}
\end{equation}
where F($t$) is the flux at time {\sl t}, A($t$) is a magnification of the source star at {\sl t}, $F_{\rm S}$ and $F_{\rm b}$ are baseline fluxes from the source and blend stars, respectively.

In this section, we estimate limb darkening and search for the best-fit model. Then we investigate the parallax effect and other models.

\subsection{Limb Darkening}
In this event, the caustic entry and exit were observed. This allows an estimate of the finite source star parameter, $\rho \equiv \theta_{\rm *}/\theta_{\rm E}$. To get $\rho$ properly, we have to apply limb darkening in the finite source calculation. We adopt a linear limb-darkening surface brightness, $S_{\rm \lambda}(\vartheta)$, given by,
\begin{equation}
S_{\rm \lambda}(\vartheta) = S_{\rm \lambda} (0) [1 - u_{\rm \lambda}(1 - \cos(\vartheta))]
\end{equation}
where $\vartheta$ is the angle between the normal to the stellar surface and the line of sight, and $u_{\rm \lambda}$ is a limb darkening coefficient in each filter. The effective temperature calculated from the source color discussed in section 5 is $T_{\rm eff} \sim 4930 \ {\rm K}$ \citep{gon09}. Assuming $T_{\rm eff} = 5000 \ {\rm K}$, surface gravity $\log g = 4.5 \ {\rm cm \ {s}^{-2}}$ and metallicity $\log [M/H] = 0$, the limb darkening coefficients  selected from Claret (2000) are $u_{\sl I} = 0.5880$, $u_{\sl V} = 0.7592$ and $u_{\rm Red} = 0.6809$ for each filter.

\subsection{Best-fit Model}
We search for the best-fit model over a wide range of values of microlensing parameters by using the Markov Chain Monte Carlo (MCMC) algorithm \citep{ver03} and the image-centered ray-shooting method \citep{ben96, ben10}. To find the global best model, we first conduct a grid search by fixing three parameters, $q$, $s$ and $\alpha$, at 20000 different grid points with other parameters free. Next, by using  the best 100 smallest $\chi^2$ models as a initial parameters, we search for the best-fit model by refining all parameters. 

\begin{figure*}
\begin{center}
\includegraphics[scale=.60,angle=270]{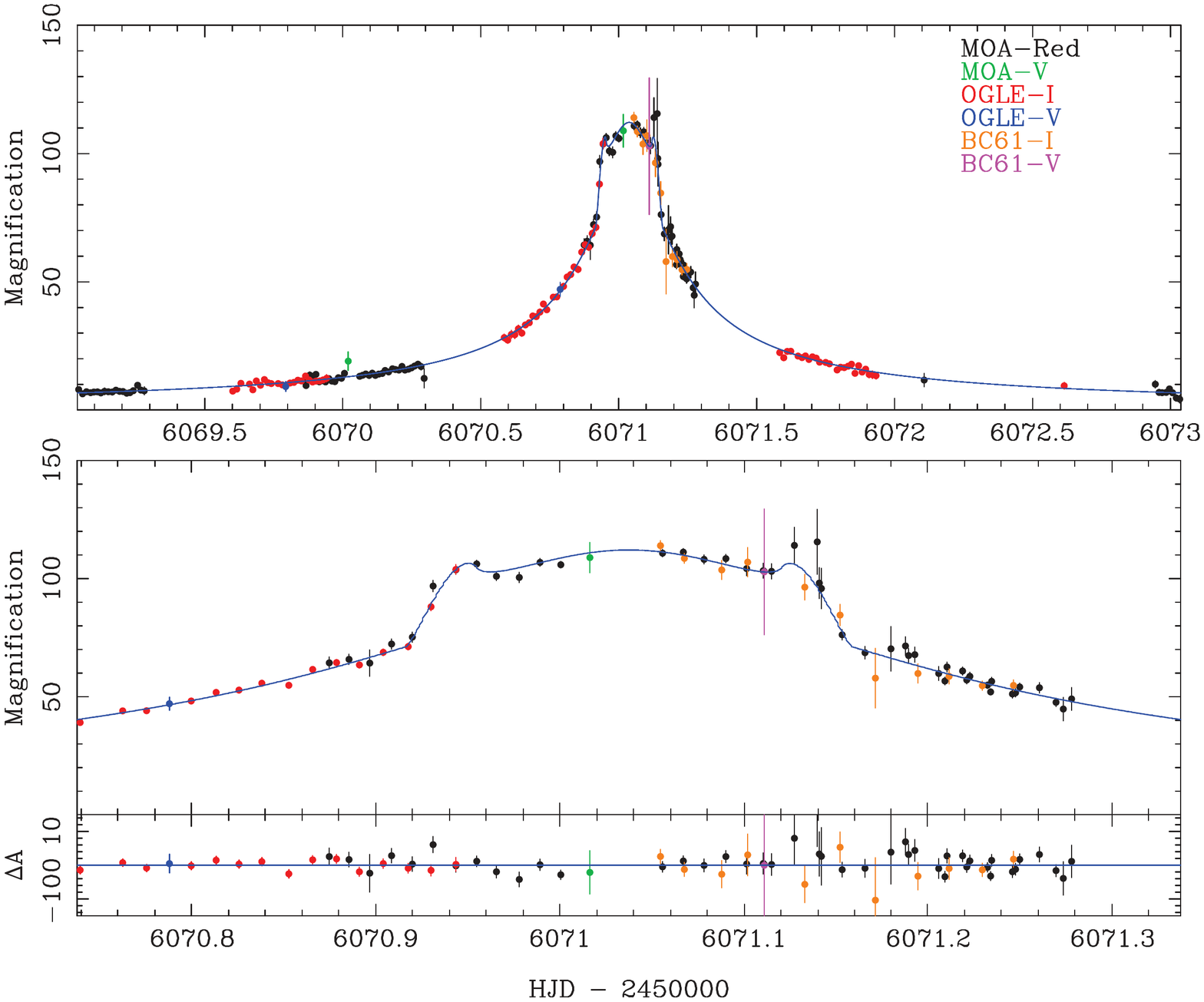}
\caption{Light curve of OGLE-2012-BLG-0724/MOA-2012-BLG-323. The top panel shows the magnified part of the light curve, the middle panel shows a close up of the anomaly and the bottom panel shows the residual from the best model. The data points of B\&C-{\sl I} and B\&C-{\sl V} are binned by 0.02 day for display purposes. Models have been fitted to unbinned data.\label{fig1}}
\end{center}
\end{figure*}

\begin{figure*}
\begin{minipage}{0.5\hsize}
\begin{center}
\includegraphics[scale=.34]{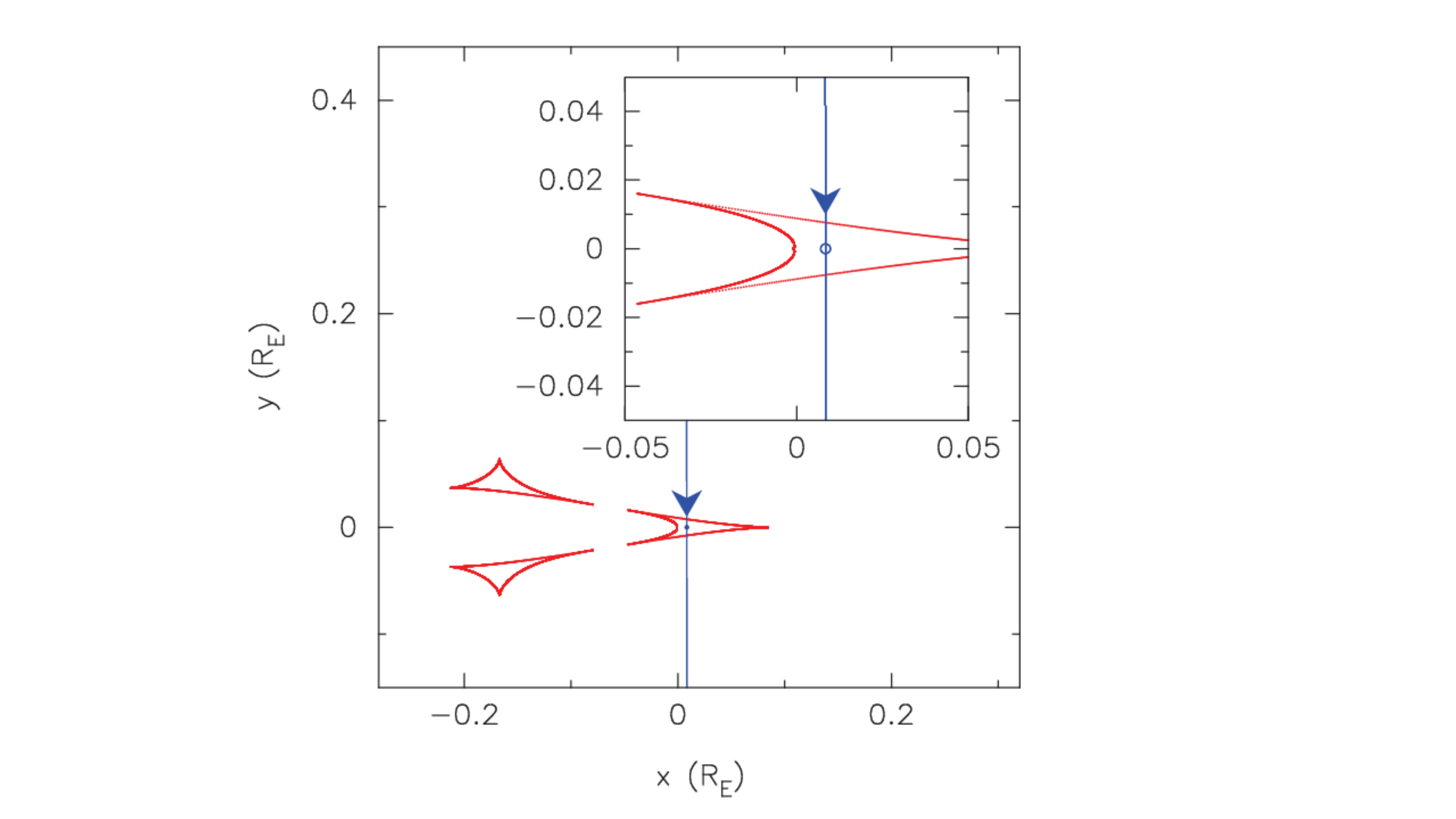}
\end{center}
\end{minipage}
\begin{minipage}{0.5\hsize}
\begin{center}
\includegraphics[scale=.34]{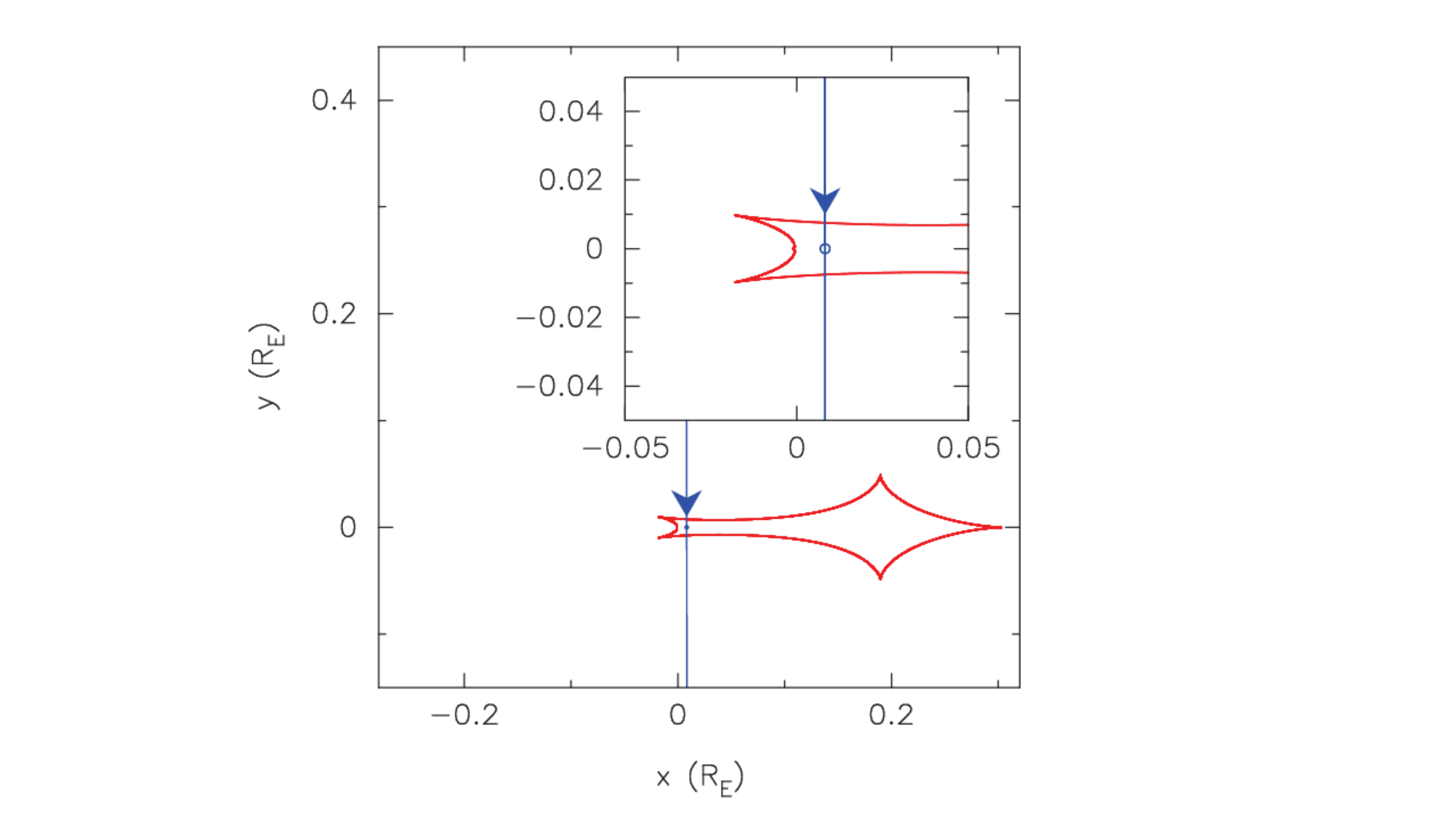}
\end{center}
\end{minipage}
\caption{Caustic geometries for both close (left) and wide (right) models are shown by the red curves. The blue lines show the source trajectory with respect to the lens system, with arrows indicating the direction of motion. The small blue circles on the lines indicate the size of the source star.\label{fig2}}
\end{figure*}

The best-fit light curve and the parameters are shown in Figure \ref{fig1} and Table \ref{tab2}. The $\Delta\chi^2$ of these models compared to the PSPL and FSPL (finite-source point-lens) are about 373 and 210, respectively, so the planetary signal is detected confidently. As is often the case for high magnification event, there are two degenerate solutions with a separation of $s$ and $1/s$ that can explain the light curve, which is known as the "close-wide" degeneracy. We call the model with $s<1$ and $s>1$ as the close and wide models, respectively. The other parameters are almost the same. These solutions are equally preferred with  $\Delta\chi^2$ of 0.2. The overall shape of the caustics of these two models are different  but the central part where the source crosses are very similar, see Figure \ref{fig2}. Thus we cannot distinguish between the models.

\subsection{Parallax Model}
Microlensing parallax is an effect where the orbital motion of the Earth causes the apparent lens-source relative motion to deviate from a constant velocity. This can be described by the microlensing parallax vector \mbox{\boldmath $\pi$}$_{\rm E}=(\pi_{\rm E,N}, \pi_{\rm E,E})$ whose direction is the direction of the lens-source relative motion projected on the sky and its amplitude,  $\pi_{\rm E}={\rm AU}/\tilde{r}_{\rm E}$, is the inverse of the Einstein radius, projected to the observer plane.

If the parallax effect and finite source effect are measured in a gravitational microlensing event, we can calculate the lens properties uniquely by assuming the distance to the source star, $D_{\rm s}$, as
$M_{\rm L}=\theta_{\rm E}/(\kappa\pi_{\rm E})$ and 
$D_{\rm L}={\rm AU}/({\pi_{\rm E}\theta_{\rm E}+\pi_{\rm s}})$
where $\kappa=4G/(c^2{\rm AU})=8.144 \ {\rm mas} \ M^{-1}_{\odot}$ and $\pi_{\rm s}={\rm AU}/D_{\rm s}$ \citep{gou00}.

\begin{figure*}
\begin{center}
\includegraphics[scale=.60]{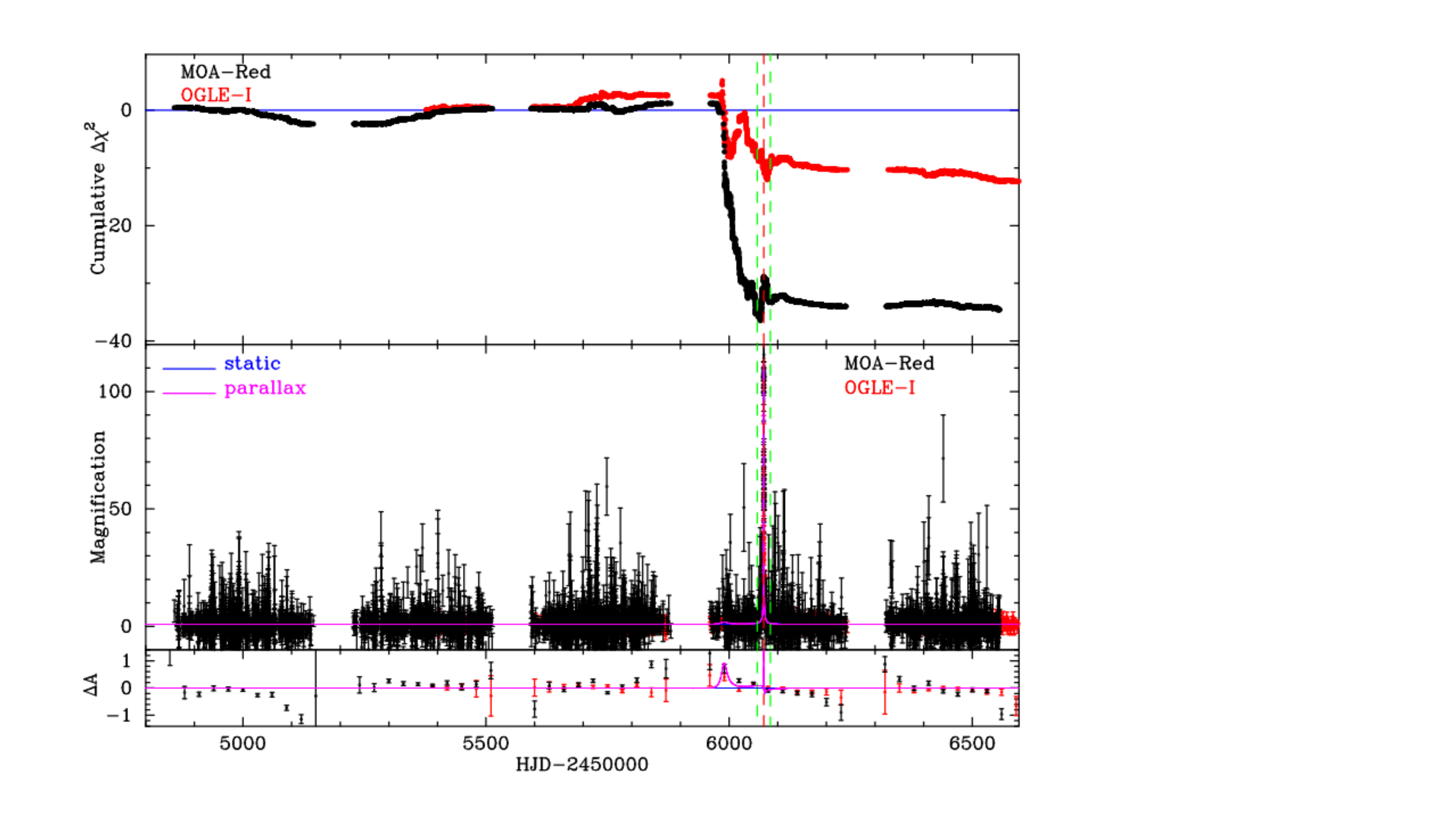}
\caption{The difference in the cumulative $\chi^2$ between the best static model and the model with the parallax effect are shown on the top panel. The middle and bottom panels show the light curves of the static model (blue) and the parallax model (magenta), and the residuals from the static model, respectively. The data in the bottom panel are binned by 30 days for clarity. The period between the dashed green lines indicates the main magnified part of this event.\label{fig8}}
\end{center}
\end{figure*}

We searched for the best model with the parallax effect and found a model with an improvement in $\chi^2$ of $\Delta\chi^2=\chi^2_{\rm stat} - \chi^2_{\rm para} \sim 47$. However the value of the parallax parameters are $\pi_{\rm E,E}=1.53\pm0.18$ and $\pi_{\rm E,N}=-8.00\pm0.63$, which are much larger than typical values, i.e., unity or less. If they were real, then the lens host star would be a close to three Jupiter-masses. Figure \ref{fig8} shows the difference in the cumulative $\chi^2$ between the static model and the model with the parallax effect. The $\Delta\chi^2$ of MOA-Red and OGLE-{\sl I} are $\Delta\chi^2_{\rm MOA- Red}\sim35$ and $\Delta\chi^2_{{\rm OGLE-}{\sl I}}\sim12$, respectively. The largest improvement in $\chi^2$ due to parallax comes from very low level continuous deviations around $HJD' \sim 6000$ which is far from the main magnified part of the event. This period corresponds to the beginning of the observation season and the GB fields were observed at low altitudes. It is known that the light curves taken at high airmass are affected by the systematics due to the basic parameters of the static model, at both the beginning and the end of the season as shown in the bottom panel of Figure \ref{fig8}. For this reason, we regard this parallax signal as not real, but due to residual systematics which could not be corrected by the de-trending process applied to the light curve. Therefore, we decided not to include the parallax effects in the following analysis. 

\subsection{Search for Degenerate Solution}
\begin{figure}
\includegraphics[scale=.70,angle=270]{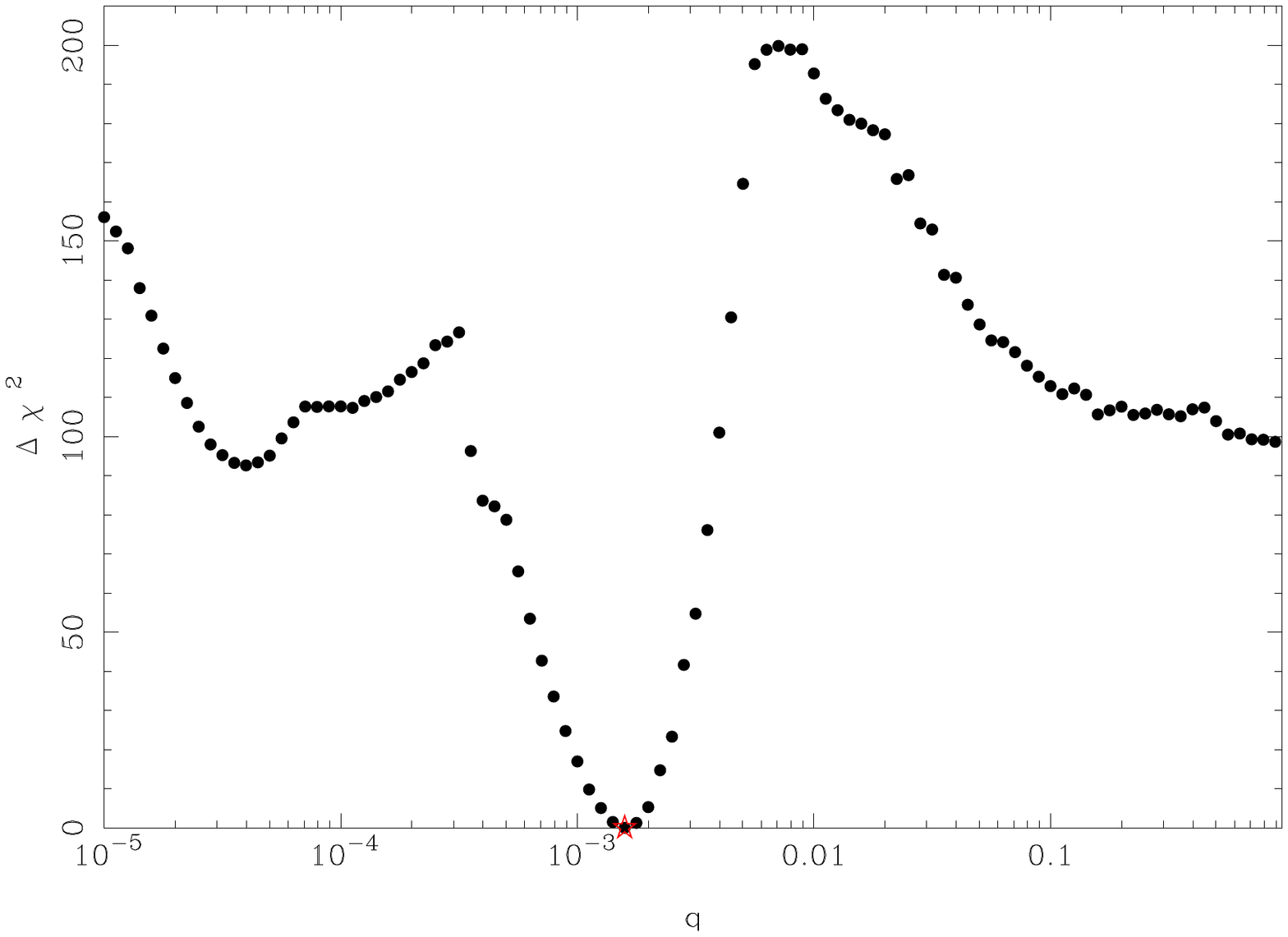}
\caption{$\Delta\chi^2$ from the best-fit model as a function of $q$. \label{fig3}}
\end{figure}

In the initial grid search for best-fitting models, we found some local minima with stellar binary mass ratios. To check the uniqueness of the best planetary model, we inspected the models with the mass ratio $q$ in the range of $-5 < \log q < 0$ carefully. Figure \ref{fig3} shows $q$ versus $\Delta\chi^2$. We found two local minima around $q \sim 4\times10^{-5} \ {\rm and} \sim 1$, which are planetary mass ratio and binary mass ratio, respectively. But the $\Delta\chi^2$ compared from the best-fit models are about 91 and 100, respectively, so we conclude the best models are superior by about 9$\sigma$ and 10$\sigma$.

\section{LENS PROPERTY}
The lens physical parameters can not be derived directly because a credible parallax effect is not measured in this event. 
Only the angular Einstein radius $\theta_{\rm E}=\theta_{\rm *}/\rho$, can be measured from the finite source effect, where $\theta_{\rm *}$ is the angular source star radius. 

\begin{figure}
\begin{center}
\includegraphics[scale=.70,angle=270]{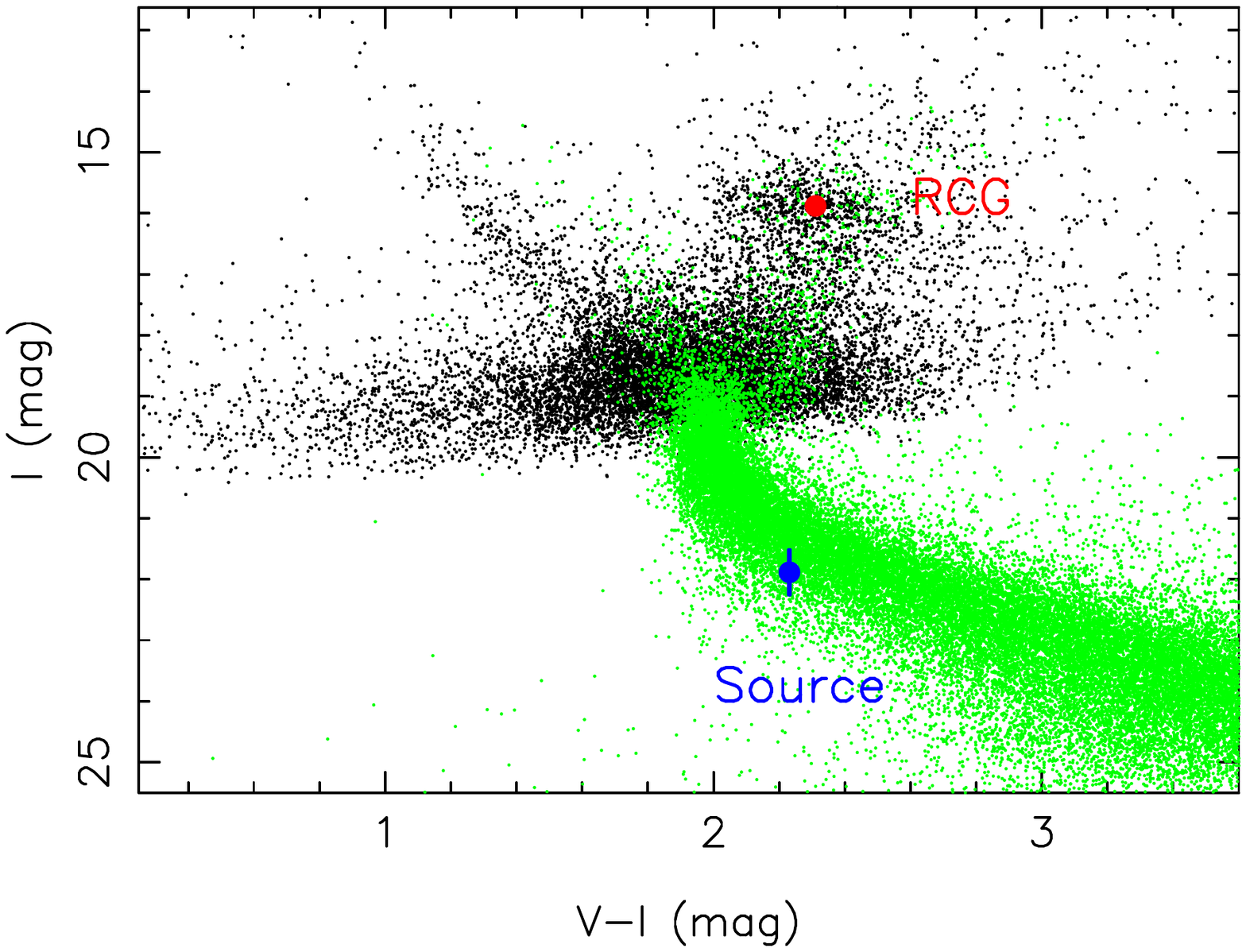}
\caption{Color-magnitude diagram (CMD) of the stars within $2'$ of OGLE-2012-BLG-0724 is shown as black dots. The green dots show the {\sl HST} CMD of Holzman et al. (1998) whose extinction is adjusted to match the OGLE-2012-BLG-0724 by using the Holtzman field red clump giant (RCG) centroid of $({\sl V-I,I})_{{\rm RCG,}{\sl HST}} = (1.62,15.15)$ (Bennett et al. 2008).  The filled red and blue circles indicate the center of the RCG and the source color and magnitude, respectively. \label{fig4}}
\end{center}
\end{figure}

Figure \ref{fig4} shows the OGLE instrumental $({\sl V-I, I})$ CMD within $2'$ around the source star. The centroid of Red Clump Giants (RCG), $({\sl V-I,I})_{\rm RCG} = (2.31,15.88) \pm (0.01,0.02)$ and the best fit source color and magnitude, $({\sl V-I, I})_{\rm S}=(2.231,21.89) \pm (0.002,0.07)$ are shown as filled red and blue circles, respectively. In Figure \ref{fig4}, we also show the stars in Baade's window observed by the {\sl HST} \citep{hol98}, which are corrected for the extinction and reddening by the RCG position in the {\sl HST} CMD, $({\sl V-I,I})_{{\rm RCG,}{\sl HST}}=(1.62,15.15)$ \citep{ben08}. They are plotted as green dots in Figure \ref{fig4}. We {\bf find out} that the best fit source color and magnitude are consistent with bulge main sequence stars.  Assuming the source suffers the same dust extinction and reddening as the RCGs and using the expected extinction-free RCG centroid $({\sl V-I, I})_{\rm RCG,0} = (1.06, 14.42) \pm (0.06, 0.04)$ at this position \citep{ben13,nat13}, we estimated the extinction-free color and magnitude of the source as $({\sl V-I,I})_{\rm S,0}=(0.98,20.43) \pm (0.06,0.09)$. We also independently derived the source color and magnitude from the MOA-Red and MOA-{\sl V} bands which are transformed to standard {\sl I} and {\sl V}-bands, and the CMD within $5'$ around the source star in the MOA reference image, and confirmed the above results. We adopt the color from the OGLE data as it is more accurate.

The angular source radius $\theta_{\rm *}$ is calculated by using the observed $(V-I, I)_{\rm S,0}$ and the relation between the limb-darkened stellar angular diameter, $\theta_{\rm LD}$, $(V-I)$ and {\sl I} from \citet{boy14} \& \citet{fuk15},
\begin{equation}
\log \theta_{\rm LD} = 0.5014 + 0.4197(V - I) - 0.2I.
\end{equation}
This gives $\theta_{\rm *} \equiv \theta_{\rm LD}/2 = 0.336 \pm 0.025 \ {\rm \mu as}$, whose error includes the $\sigma_{\sl (V-I)}, \sigma_{\sl I}$ and the 2\% uncertainty in Equation 4.

We derived the angular Einstein radius $\theta_{\rm E}$ and the geocentric lens-source relative proper motion $\mu_{\rm geo}$ for close and wide models as follows,
\begin{equation}
\theta_{\rm E} = \frac{\theta_*}{\rho} = 0.239 \pm 0.028 \ {\rm mas \ (close)} 
\end{equation}
\begin{equation}
 = 0.242 \pm 0.029 \ {\rm mas \ (wide)},
\end{equation}
\begin{equation}
\mu_{\rm geo} = \frac{\theta_{\rm E}}{t_{\rm E}} = 6.52 \pm 0.87 \ {\rm mas \ yr^{-1} \ (close)} 
\end{equation}
\begin{equation}
 = 6.55 \pm 0.90 \ {\rm mas \ yr^{-1} \ (wide)} .
\end{equation}

\begin{figure*}
\begin{center}
\includegraphics[scale=.60,angle=270]{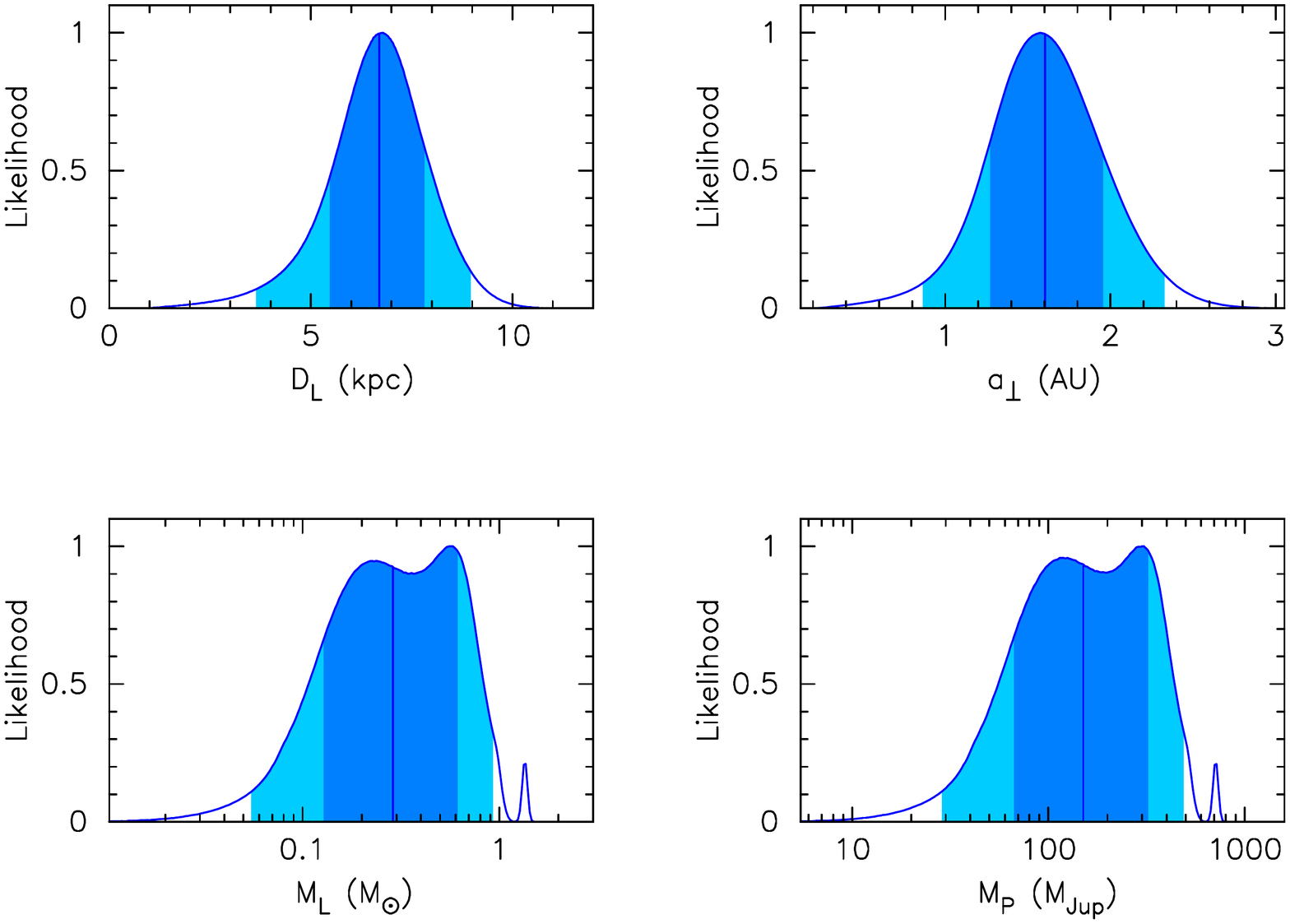}
\caption{Probability distribution of lens properties from the Bayesian analysis. The dark and light blue regions indicate the 68.3\% and 95.4\% confidence interval, respectively, and the vertical blue lines indicate the median value. \label{fig5}}
\end{center}
\end{figure*}

\begin{figure*}
\begin{center}
\includegraphics[scale=.60,angle=270]{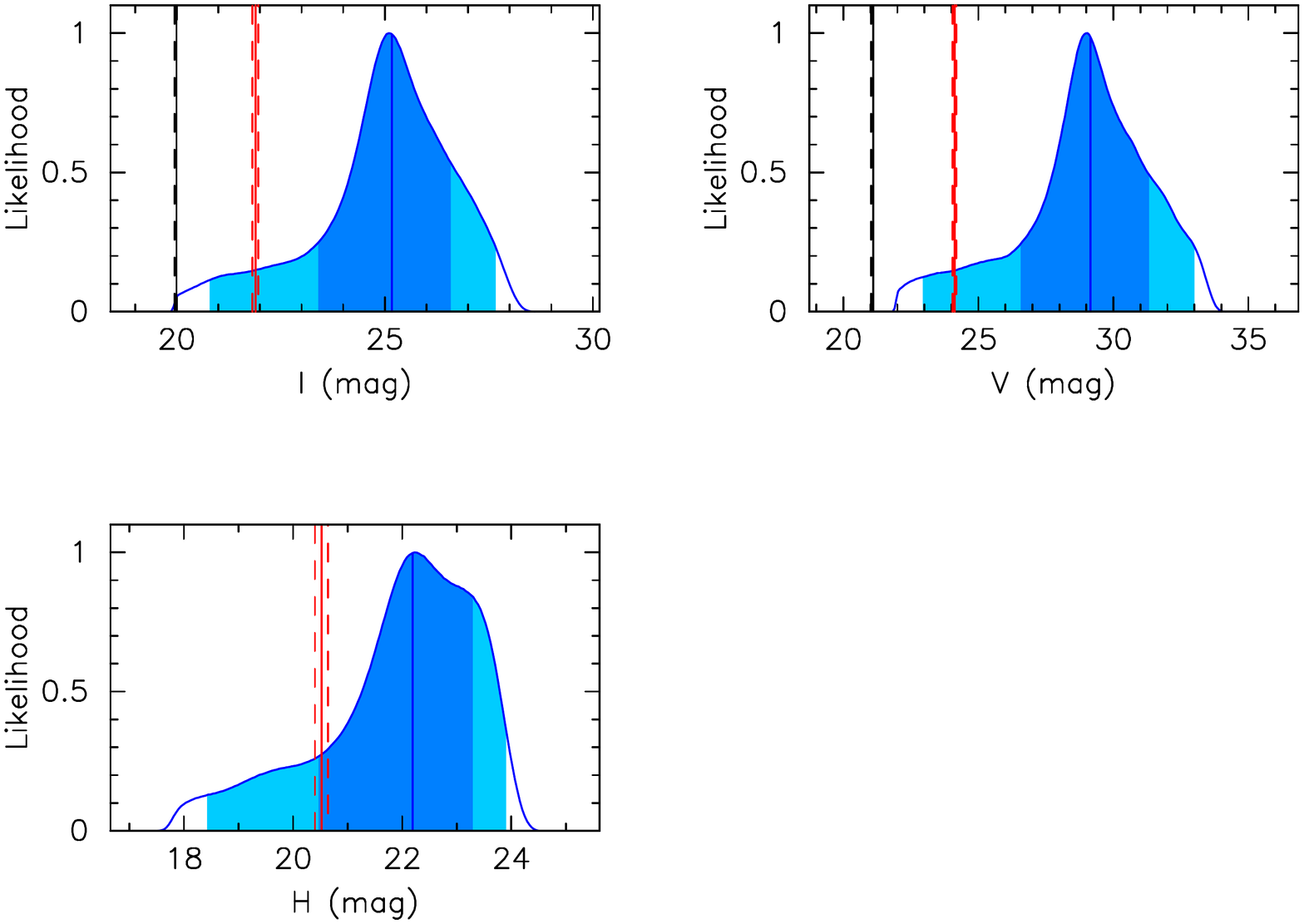}
\caption{Probability distribution of {\sl I}-, {\sl V}- and {\sl H}-band magnitudes for the lens star with the extinction from a Bayesian analysis. The dark and light blue regions indicate the 68.3\% and  95.4\% confidence interval, respectively, and the vertical blue lines indicate the median parameter value. The black vertical lines indicate the upper limit of the {\sl I}- and {\sl V}-band lens magnitudes with the extinction derived from the blending flux. The red solid lines show the source star magnitudes with the extinction and the red dashed lines are their 1$\sigma$ errors. The {\sl I}- and {\sl V}-band source magnitudes are derived from the light curve fitting and the {\sl H}-band source magnitude is estimated from the the stellar color-color relation in \citet{ken95} and the extinction estimated from the extinction law by \citet{car89}.}\label{fig6}
\end{center}
\end{figure*}

We conduct a Bayesian analysis to get the probability distribution of the lens properties \citep{bea06,gou06,ben08}.  We calculate the likelihood by using the Galactic model \citep{han03} and the observed $t_{\rm E}$ and $\theta_{\rm E}$. We also used the de-reddened blending flux which includes the lens and unrelated stars derived from the OGLE-{\sl I} and OGLE-{\sl V} light curves as the upper limit for the lens brightness,
\begin{equation}
I_{\rm b,0} = 18.54 \pm 0.05,
\end{equation}
\begin{equation}
V_{\rm b,0} = 18.39 \pm 0.08,
\end{equation}
Because the $\chi^2$ and physical parameters of close and wide models are similar, we combined the probability distribution of these models by weighting the probability distribution of the wide model by $e^{-\Delta\chi^2/2}$, where $\Delta\chi^2=\chi^2_{\rm wide}-\chi^2_{\rm close} \sim 0.20$. Figure \ref{fig5} and \ref{fig6} show the probability distribution of the lens properties derived from the Bayesian analysis. According to the results, the lens host star is an M-dwarf with a mass of $M_{\rm L}=0.29_{-0.16}^{+0.33} \ M_{\sun}$ and its distance is $D_{\rm L}=6.7_{-1.2}^{+1.1} \ {\rm kpc}$ away from the Earth, and the mass of the planet is $m_{\rm P}=0.47_{-0.26}^{+0.54} \ M_{\rm Jup}$ and the projected separation is $a_{\perp}=1.6_{-0.3}^{+0.4} \ {\rm AU}$. If we assume a circular and randomly oriented orbit for the planet, the 3-dimensional semi-major axis is expected to be $a=2.0_{-0.5}^{+1.0} \ {\rm AU}$. The probability distribution of {\sl I}-, {\sl V}- and {\sl H}-band magnitudes of the lens star with the extinction is shown in Figure \ref{fig6}. The extinction for the lens is assumed to be the same as that for the source because the lens is predicted to be near the Galactic bulge. The {\sl H}-band magnitude of the source with the extinction is calculated to be $H_{\rm S}=20.52 \pm 0.12$ from the extinction-free $(I-H)$ color of the source estimated from $(V-I)$ color using the stellar color-color relation \citep{ken95} and the {\sl H}-band extinction $A_H$ estimated from $A_I$ and $A_V$ using the extinction law by \citet{car89}. The lens is predicted to be fainter than the source by $\sim 3$ magnitude in the {\sl I}-band, $\sim 5$ magnitude in the {\sl V}-band and $\sim 2$ magnitude in the {\sl H}-band, respectively. Thus it is not easy to detect the lens flux by future high resolution follow-up observations.  

\section{DISCUSSION AND CONCLUSION}
Originally, OGLE-2012-BLG-0724Lb was thought to comprise a binary star lens system. In this work, we concluded that OGLE-2012-BLG-0724Lb is a Saturn mass planet orbiting around an M-dwarf. This system is of a type typically found by gravitational microlensing. Recently, \citet{fuk15} found a possible pileup of sub-Jupiters ($0.2 < m_{\rm P}/M_{\rm Jup} < 1$) in contrast to a lack of Jupiters ($\sim 1-2 \ M_{\rm Jup}$) around M-dwarfs, suggesting that Jupiter-mass planets are rarely formed around M-dwarfs, as predicted by core accretion models (e.g. \citet{ida05}). OGLE-2012-BLG-0724Lb belongs to the sub-Jupiter class, supporting the above idea. However, the statistical error of the abundance of these planets is still large. A larger sample will give us a better understanding of the planetary formation mechanisms.

In statistical analysis leading to estimates of planet abundances, we need to beware events which have degenerate solutions. For example, OGLE-2008-BLG-355Lb was originally published as a stellar binary, but the reanalysis of \citet{kos14} concluded that it was a planet. In such cases, we need to investigate the wide range of parameters so as not to be fooled by local minima. Event OGLE-2012-BLG-0724 has a relatively short time scale, the anomaly lasted for only 6 hrs, and it was not very bright even at the high-magnification peak, which was difficult for follow-up observations by small telescopes. We could figure out the planetary nature and its superiority over the binary solution thanks to the high-cadence observations by OGLE and MOA. 

The recent high-cadence survey observations by medium-aperture telescopes equipped with a wide field-of-view camera as used by the MOA, OGLE and KMTNet groups are going to detect many planetary events without dense follow-up observations. But if we seek more rare events like an Earth-mass planetary event, the sampling during short anomalies will be limited and we will face similar marginal conclusions. This is also true at the boundary of detection limits even for the future space based observations by {\sl WFIRST} \citep{spe13} and {\sl Euclid} \citep{pen13}. This work is a test case for analysis efforts in such marginal events.

We could not measure the mass of the system because microlensing parallax was not detected. We can measure the lens properties if we can directly resolve the lens flux in high resolution follow-up observations by the {\sl HST} or large ground telescopes, such as the Keck, VLT and Subaru \citep{ben06, ben15, bat15}. The unrelated blending stars can usually be separated from the lens and source by high resolution follow-up observations. The lens-source relative proper motion is $\mu_{\rm geo}=6.52 \pm 0.87 \ {\rm mas \ yr^{-1}}$. This implies that the source and lens will be separated by $\sim 1$ HST pixel ($\sim 39.12 \pm 5.22$ mas) by 2018. But in the {\sl I}-band, the median lens magnitude is predicted to be very faint compared to the source magnitude by $\sim 3$ magnitude (see Figure 7) and it makes difficult to observe the lens star. From Figure 7, we see that the {\sl H}-band magnitude of the median prediction for the lens and source differs by $\sim 2$ magnitude. The lens is likely to be too faint to be detected until the lens and source are resolved. The future high resolution images by the {\sl HST}, {\sl JWST} or ground based telescopes using Adaptive Optics such as TMT in more than 10 years after peak magnification, which is in 2022, may detect the lens host star and determine the lens properties uniquely.

However, if the lens and source have different color, the apparent centroid shift of the lens plus source differs from images in different filter bands. In OGLE-2003-BLG-235, the color-dependent centroid offset was detected at the level of 0.6 mas by the {\sl HST} \citep{ben06}. In OGLE-2012-BLG-0724, we expect to detect the color-dependent centroid shift of ${\rm d}x=1.0$ mas in ${\rm d}t= 4$ years after the peak magnification from the relation ${\rm d}x = {\rm d}t \times (f_I -f_V ) \times \mu_{\rm geo}$. Here $\mu_{\rm geo} = 6.52 \ {\rm mas \ yr^{-1}}$ and the fraction of the lens+source flux that is due to the lens, $f_{\rm lens}$, is $f_I = 0.05$ and $f_V = 0.01$ in the {\sl I}- and {\sl V}-band, respectively. Thus the follow-up observations in different filter bands by the {\sl HST} in 2016 can detect the lens flux and determine the lens properties.
\\
\par
TS acknowledges the financial support from the JSPS, JSPS23103002, JSPS24253004 and JSPS26247023. The MOA project is supported by the grant JSPS25103508 and 23340064. The OGLE project has received funding from the National Science Centre, Poland, grant MAESTRO 2014/14/A/ST9/00121 to AU. OGLE Team thanks Profs. M. Kubiak and G. Pietrzy{\'n}ski, former members of the OGLE team, for their contribution to the collection of the OGLE photometric data over the past years. DPB acknowledges support from NSF grants AST-1009621 and AST-1211875, as well as NASA grants NNX12AF54G and NNX13AF64G. Work by IAB was supported by the Marsden Fund of the Royal Society of New Zealand, contract no. MAU1104. NJR is a Royal Society of New Zealand Rutherford Discovery Fellow. AS, ML and MD acknowledge support from the Royal Society of New Zealand. AS is a University of Auckland Doctoral Scholar. NK is supported by Grant-in-Aid for JSPS Fellows.

\newpage

\begin{table}
\begin{center}
\caption{The Data Sets Used to Model the OGLE-2012-BLG-0724 Light Curve and the Error Correction Parameters\label{tab1}}
\begin{tabular}{lcccc} \toprule
Data set &  $k$ & $e_{\rm min}$ & Number of Data \\ \midrule
MOA-Red  & 0.906 & 0.011 & 27964 \\
MOA-{\sl V}  & 0.832 & 0 & 81 \\
OGLE-{\sl I}  &1.456 & 0.009 & 7449 \\
OGLE-{\sl V}  &1.603 & 0 & 108 \\
B$\&$C-{\sl I}  &0.772 & 0 & 171 \\
B$\&$C-{\sl V}  & 0.445 & 0 & 35 \\ \bottomrule
\end{tabular}
\end{center}
\end{table}

\begin{table}
\begin{center}
\caption{The Best-Fit Model Parameters For Both the Close and Wide Models \label{tab2}}
\begin{tabular}{lcc} \toprule
parameter　& close & wide \\ 
1$\sigma$ error & (s$<$1) & (s$>$1) \\ \midrule
$t_0$(HJD') & 6071.0381 & 6071.0381 \\
 & 0.0010 & 0.0010 \\ 
$t_{\rm E}$(days) & 13.3556 & 13.4935 \\
 & 0.8479 & 0.8639 \\
$u_0$($10^{-2}$) & 0.8383 & 0.8268 \\
 & 0.0612 & 0.0611 \\
$q$($10^{-3}$) & 1.5848 & 1.5935 \\
 & 0.1451 & 0.1617 \\
$s$ & 0.9195 & 1.0990 \\
 & 0.0053 & 0.0063 \\
$\alpha$(radian) & 1.5700 & 1.5696 \\
 & 0.0117 & 0.0123 \\
$\rho$($10^{-3}$) & 1.4098 & 1.3922 \\
 & 0.1282 & 0.1338 \\
$\chi^2$ & 35752.40 & 35752.60 \\
d.o.f & 35788 & 35788 \\ \bottomrule
\end{tabular}
\end{center}
\end{table}



\begin{thebibliography}{dummy}
\bibitem[Alard \& Lupton(1998)]{ala98} Alard, C., \& Lupton, R. H. 1998, ApJ, 503, 325
\bibitem[Batista et al.(2015)]{bat15} Batista, V., Beaulieu, J, -P., Bennett, D. P., et al. 2015, ApJ, 808, 170
\bibitem[Beaulieu et al.(2006)]{bea06} Beaulieu, J.-P., Bennett, D. P., Fouqu\'e, P., et al. 2006, Natur, 439, 437
\bibitem[Bennett(2010)]{ben10} Bennett, D. P. 2010, ApJ, 716, 1408
\bibitem[Bennett et al.(2006)]{ben06} Bennett, D. P. Anderson, J., Bond, I. A. 2006, ApJ, 647, L171
\bibitem[Bennett et al.(2008)]{ben08} Bennett, D. P., Bond, I. A., Udalski, A., et al. 2008, ApJ, 684, 663
\bibitem[Bennett et al.(2015)]{ben15} Bennett, D. P., Bhattacharya, A., Anderson, J., et al. 2015, ApJ, 808, 169
\bibitem[Bennett \& Rhie(1996)]{ben96} Bennett, D. P., \& Rhie, S. H. 1996, ApJ, 472, 660
\bibitem[Bensby et al.(2013)]{ben13} Bensy, T., Yee, J. C., Felzing, S., et al. 2013, A\&A, 549, A147
\bibitem[Bond et al.(2001)]{bon01} Bond, I. A., Abe, F., Dodd, R. J., et al. 2001, MNRAS, 327, 868
\bibitem[Borucki et al.(2011)]{bor11} Borucki, W. J., Koch, D. G., Basri, G., et al. 2011, ApJ, 736, 19
\bibitem[Boyajian et al.(2014)]{boy14} Boyajian, T. S., van Belle, G., \& von Braun, K. 2014, AJ, 147, 47 
\bibitem[Butler et al.(2006)]{but06} Butler, R. P., Wright, J. T., Marcy, G. W., et al. 2006, ApJ, 646, 505 
\bibitem[Cardelli et al.(1989)]{car89} Cardelli, J. A., Clayton, G. C., \& Mathis, J. A., 1989, ApJ, 345, 245C
\bibitem[Cassan et al.(2012)]{cas12} Cassan, A., Kubas, D., Beaulieu, J.\_P., et al. 2012, Natur, 481, 167
\bibitem[Cassan \& Ranc(2016)]{cas16} Cassan, A., \& Ranc, C. 2016, MNRAS, 458, 2074
\bibitem[Clanton \& Gaudi(2014)]{cla14} Clanton, C., \& Gaudi, B. S. 2014, ApJ, 791, 90
\bibitem[Claret(2000)]{cla00} Claret, A. 2000, A\&A, 363, 1081
\bibitem[Fukui et al.(2015)]{fuk15} Fukui, A., Gould, A., Sumi, T., et al. 2015, ApJ, 809, 74
\bibitem[Gonz\'alez \& Bonifacio(2009)]{gon09} Gonz\'alez Hern\'andez, J. I., \& Bonifacio, P., et al. 2009, A\&A, 497, 497
\bibitem[Gould \& Loeb(1992)]{gou92} Gould, A., \& Loeb, A. 1992, ApJ, 396, 104
\bibitem[Gould(2000)]{gou00} Gould, A. 2000, ApJ, 5442, 785
\bibitem[Gould et al.(2006)]{gou06} Gould, A. 2006, ApJ, 644, L37
\bibitem[Gould et al.(2010)]{gou10} Gould, A., Dong, S., Gaudi, B. S., et al. 2010, ApJ, 720, 1073
\bibitem[Gould \& Horne(2013)]{gou13} Gould, A., \& Horne, K. 2013, ApJL, 779, L28
\bibitem[Gould \& Yee(2014)]{gou14} Gould, A., \& Yee, J. C., 2014, ApJ, 784, 64
\bibitem[Han \& Gould(2003)]{han03} Han, C., \& Gould, A. 2003, ApJ, 592, 172
\bibitem[Holtzman et al.(1998)]{hol98} Holtzman, J. A., Watson, A. M., Baum, W. A., et al. 1998, AJ, 115, 1946
\bibitem[Ida \& Lin(2005)]{ida05} Ida, S., \& Lin, D. N. C. 2005, ApJ, 625, 1045
\bibitem[Kenyon \& Hartmann(1995)]{ken95} Kenyon, S. J., \& Hartmann, L., 1995, ApJSS, 101, 117
\bibitem[Kervella \& Fouqu\'e(2008)]{ker08} Kervella, P., \& Fouqu\'e, P. 2008, A\&A, 491, 855
\bibitem[Koshimoto et al.(2014)]{kos14} Koshimoto, N., Udalski, A., Sumi, T., et al. 2014, ApJ, 788, 128
\bibitem[Mayor \& Queloz(1995)]{may95} Mayor, M., \& Queloz, D. 1995, Natur, 378, 355
\bibitem[Nataf et al.(2013)]{nat13} Nataf, D. M., Gould, A., Fouqu\'e, P., et al. 2013, ApJ, 769, 88
\bibitem[Rattenbury et al.(2002)]{rat02} Rattenbury, N. J., Bond, I. A., Skuljan, J., et al. 2002, MNRAS, 335, 159
\bibitem[Penny et al.(2013)]{pen13} Penny, M. T., Kerins, E., Rattenbery, N., et al. 2013, MNRAS, 434, 2
\bibitem[Sako et al.(2008)]{sak08} Sako, T., Sekiguchi, T., Sasaki, M., et al. 2008, ExA, 22, 51
\bibitem[Schechter et al.(1993)]{sch93} Schechter, P. L., Mateo, M., \& Saha, A. 1993, PASP, 105, 1342
\bibitem[Shvartzvald et al.(2015)]{shv15} Shvartzvald, D., Maoz, D., Udalski, A., et al. 2015, MNRAS, submitted(arXiv:1510.04297)
\bibitem[Spergel et al.(2013)]{spe13} Spergel, D., Gehrels, N., Breckinridge, J., et al. 2013, arXiv:1305.5422
\bibitem[Sumi et al.(2003)]{sum03} Sumi, T., Abe, F., Bond, I. A., et al. 2003, ApJ, 591, 204
\bibitem[Sumi et al.(2010)]{sum10} Sumi, T., Bennett, D. P., Bond, I. A., et al. 2010, ApJ, 710, 1641
\bibitem[Sumi et al.(2011)]{sum11} Sumi, T., Kamiya, K., Bennett, D. P., et al. 2011, Natur, 473, 349
\bibitem[Suzuki et al.(2016)]{suz16} Suzuki, D., Bennett, D. P., Sumi, T., et al., 2016, ApJ, submitted.
\bibitem[Tomaney \& Crotts(1996)]{tom96} Tomaney, A. B., \& Crotts, A. P. S. 1996, AJ, 112, 2972
\bibitem[Udalski(2015)]{uda15a} Udalski, A., Szyma{\'n}ski, M. K., \& Szyma{\'n}ski, G. 2015, Acta Astron., 65, 1.
\bibitem[Udalski(2015)]{uda15} Udalski, A., Yee, J. C., Gould, A. 2015, ApJ, 799, 237
\bibitem[Verde \& Spergel(2003)]{ver03} Verde, L., Peiris, H. V., \& Spergel, D. N. 2003, ApJS, 148, 195
\bibitem[Zhu et al.(2015)]{zhu15} Zhu, W., Udalski, A., Gould, A. et al. 2015, ApJ, 80, 5
\bibitem[Yee et al.(2012)]{yee12} Yee, J. C., Shvartzvald, Y., Gal-Yam, A., et al. 2012, ApJ, 775, 102


\end{thebibliography}
\end{document}